\documentclass[manuscript]{aastex} 

\newcommand{\KICa}{KIC6034120} 
\newcommand{\KICb}{KIC6691930} 
\newcommand{\KICc}{KIC10528093} 
\def\vector#1{\mbox{\boldmath $#1$}}

\begin{document}
\title{Superflares on Solar-Type Stars Observed with Kepler \\ 
II. Photometric Variability of Superflare-Generating Stars : \\
A Signature of Stellar Rotation and Starspots}

\author{Yuta Notsu\altaffilmark{1},  Takuya Shibayama\altaffilmark{1}, Hiroyuki Maehara\altaffilmark{2,3}, Shota Notsu\altaffilmark{1}, Takashi Nagao\altaffilmark{1}, Satoshi Honda\altaffilmark{2,4},  Takako T. Ishii\altaffilmark{2}, Daisaku Nogami\altaffilmark{2}, and Kazunari Shibata\altaffilmark{2}}
\affil{\altaffilmark{1}Department of Astronomy, Faculty of Science, Kyoto University, 
Kitashirakawa-Oiwake-cho, Sakyo-ku, Kyoto, 606-8502, Japan}
\affil{\altaffilmark{2}Kwasan and Hida Observatories, Kyoto University, 
Kitakazan-ohmine-cho, Yamashina-ku, Kyoto, 607-8471, Japan}
\affil{\altaffilmark{3}Kiso Observatory, Institute of Astronomy, School of Science, The University of Tokyo,
10762-30, Mitake, Kiso-machi, Kiso-gun, Nagano 397-0101, Japan}
\affil{\altaffilmark{4}Center for Astronomy, University of Hyogo, 
407-2, Nishigaichi, Sayo-cho, Sayo, Hyogo, 679-5313, Japan}
\email{ynotsu@kwasan.kyoto-u.ac.jp}

\begin{abstract}
We performed simple spot-model calculations for quasi-periodic 
brightness variations of solar-type stars showing superflares, 
by using Kepler photometric data.
Most of superflare stars show quasi-periodic brightness modulations with the typical period 
of one to a few tens of days. 
Our results indicate that these brightness 
variations of superflare stars can be explained by the rotation of 
the star with fairly large starspots. 
Using the result of the period analysis, 
we investigated the relation between the energy and frequency of superflares 
and the rotation period. 
Stars with relatively slower rotation rates can still produce flares that are as energetic 
as those of more rapidly rotating stars, although the average flare frequency is lower for more slowly rotating stars.
We found that the energy of superflares are related to the total coverage of 
starspots. The correlation between the spot coverage and the flare energy in
superflares is similar to that in solar flares.
These results suggest that the energy of superflares can be explained by
the magnetic energy stored around starspots.
\end{abstract}


\section{Introduction}\label{sec:intro}
\noindent Solar flares are the most energetic explosions on the surface of the Sun,
and are thought to occur by release of magnetic energy (e.g., \citealt{shibata2011}). 
Flares are also known to occur 
on various types of stars including solar-type stars (\citealt{schaefer1989}; \citealt{Gershberg2005}). 
Among them, young stars or close binary stars sometimes produce "superflares", 
flares whose total energy is $10\sim10^6$ times more energetic ($\sim10^{33-38}$ erg) than the largest flares on the Sun ($\sim10^{32}$erg) \citep{schaefer2000}. 
Such stars generally rotate fast ($v_{\rm{rot}}\sim$a few 10 km s$^{-1}$), 
and the magnetic fields of a few kG are considered to be distributed in large regions 
on the stellar surface (\citealt{Gershberg2005}; \citealt{shibata1999}, \citeyear{shibata2002}). 
In contrast, the Sun slowly rotates ($v_{\rm{rot}}\sim$2 km s$^{-1}$), 
and the magnetic fields are weak. 
Here we define ``Sun-like" stars as solar-type stars slowly rotating 
and whose surface temperature is 5,600K $\sim$ 6,000K. 
It has been thought that superflares cannot occur on Sun-like stars. 
\\ \\
\ \  \citet{schaefer2000}, however, found 9 candidates of superflares on slowly rotating stars like the Sun.
This is extremely important in many fields including magnetic activity research in solar/stellar physics as well as the planetary habitability in astrobiology (e.g., \citealt{segura2010}).
The frequency, detailed properties, and mechanism of the superflares 
are, however, still not clear because of lack of observations. 
Therefore, it is necessary to investigate in detail how often superflares occur on solar-type stars, properties of superflares, and stellar conditions which lead to superflares. 
\\ \\ 
\ \ We have already analyzed the data by the Kepler spacecraft  \citep{koch2010}, and 
discovered 365 superflare events on 148 solar-type stars that have surface temperature of 
5,100K $\leq$ $T_{\rm{eff}} <$ 6,000K, and surface gravity of log $g \geq$ 4.0 
\citep{maehara2012}. The Kepler spacecraft  is very useful 
to detect faint brightness increases in the stellar brightness due to stellar flares 
because Kepler realized high photometric precision 
and continuous time-series data of a lot of stars over 
a long period (\citealt{walkowicz2011}; \citealt{balona2012}). 
We found that superflares whose energy is $10^{2}-10^{3}$ times larger than 
that of the most energetic flare on the Sun, 
can occur on solar-type stars once in a few thousands of years. 
Many of solar-type stars having superflares 
show quasi-periodic brightness variations
with the typical period from one day to a few tens of days. 
Such variations can be explained 
by the rotation of the star (e.g., \citealt{basri2011}; \citealt{debosscher2011}; \citealt{harrison2012}). 
\\ \\
\ \ In this paper, we investigate the brightness variations of superflare-generating stars
by considering the signature of stellar rotation and starspots. 
First we show that the brightness variation of some typical superflare-generating stars are explained by the 
rotation of a star with starspots, by simple model analyses. 
Our main purpose of the model analysis is to demonstrate that the brightness 
variations include the information of the overall areal coverage by starspots and the rotation periods of superflare stars.
Second, using the result of the period analysis, 
we consider relations between the rotation period 
and properties of superflares such as flare energy and flare frequency. 
Third, we discuss relations  between the starspot coverage and the energy of superflares 
by assuming that the brightness variations are due to the rotation.
For the Sun, it has been known that there is a positive 
correlation between the sunspot coverage  
and the energy of the largest flare observed, and that 
the energy of the largest flare depends on the magnetic energy stored 
around the sunspots (e.g., \citealt{sammis2000}).
We then investigate whether this correlation can be applied to superflare-generating stars, and 
discuss the relations between the magnetic energy of starspots and the superflare energy. 
\\ \\
\ \ 
The details of Kepler data we use in this paper are explained in Section \ref{sec:Kepler-data}. 
We show in Section \ref{sec:model-ana} the results of the spot modeling 
for some typical superflare-generating stars.
We consider the relation  between the rotation period 
and properties of superflares, on the basis of the results of the period analysis in Section \ref{sec:period-ana-plot}. 
We discuss the relation  between the starspot coverage and the energy of superflares 
by assuming that the brightness variations of superflare-generating stars are due to 
the rotation in Section \ref{sec:discussion}.
\section{Observational Data and Period Analysis}\label{sec:Kepler-data}
\noindent
We searched for flares on solar-type stars using the Kepler data
which was taken during the period from April 2009 to September 2010 (quarter 0$\sim$6). 
These data were retrieved from the Multimission Archive at Space 
Telescope Science Institute (MAST)\footnote{http://archive.stsci.edu/kepler/}. 
We used the effective temperature ($T_{\rm{eff}}$) and the surface 
gravity ($\log g$) in the Kepler Input Catalog \citep{brown2011} to select solar-type 
stars. The selection criteria are as follows; $5,100 \le T_{\rm{eff}} < 6,000$, 
and $\log g \ge 4.0$. The total number of solar-type stars is about 90,000. 
The length of the observation period during each quarter and the number of 
solar-type stars are summarized in Table \ref{data_summary}.
\\ \\
\ \ 
We analyzed the long-cadence (time resolution of about 30 min) flux 
detrended by PDC-MAP pipeline \citep{stumpe2012} for detection of superflares. 
Details on the detection 
method are described in \citet{maehara2012} and \citet{shibayama2013}. 
The version of the data we use in this paper are shown in Table \ref{data_summary}.
\\ 
\\
\ \ Figure \ref{lc} shows the typical light curves of solar-type stars 
having superflares (\KICa \ \& \KICb, Quarter 2; \KICc, Quarter 0-2). 
\KICa \ shows a lightcurve of a simple sine-like profile which can be well reproduced with one large starspot.
The lightcurve of \KICb \ requires two starspots to explain the shoulder of the rising part. 
\KICc \ also needs two starspots, and its long-term amplitude variation cannot be explained by the solid rotation model. 
These parts of the data are selected to make these features clearly demonstrated. 
Light curves of all solar-type stars having superflares 
reported in \citet{shibayama2013} are shown in Online-only Figures. 
The period of the brightness variation was estimated by the discrete 
Fourier transform (DFT) method. 
We carried out the following 5 procedures: 
(1) We measured the standard deviations ($\sigma$) of the data in each quarter, and deleted all the data of a quarter whose $\sigma$ value is over three times larger than the mean value.
(2) The linear trend was removed in the brightness during each quarter. 
(3) The gaps in the mean brightness between each quarter were adjusted.
(4) We took the power spectra of the period from 0.1 day to 45 day, since the 
reduction pipeline seems to generate spurious peaks at timescales longer than 45 
days, a half of typical lengths of each quarter\footnote{For the detrend method by the Kepler pipeline, 
see \\ http://keplergo.arc.nasa.gov/PyKEprimerCBVs.shtml}. 
(5) We chose the peak in the power spectrum whose amplitude has the highest ratio 
to the red noise spectrum (e.g., \citealt{press1978}; \citealt{vaughan2005}) as the period 
of the brightness variation of the star, 
since in some cases simply taking the highest peak can lead to choose spurious peaks at long timescales.
Figure \ref{period_ana} is the results of the period analysis for the brightness variation shown in Figure \ref{lc}. 
Such stars show quasi-periodic brightness variations 
with the typical period from one day to a few tens of days. 
\\ 
\section{Model Calculation of Brightness Variation of Rotating Stars with Large Starspots}\label{sec:model-ana}
\begin{figure}[htpb]
\begin{center}
\epsscale{0.7}
\plotone{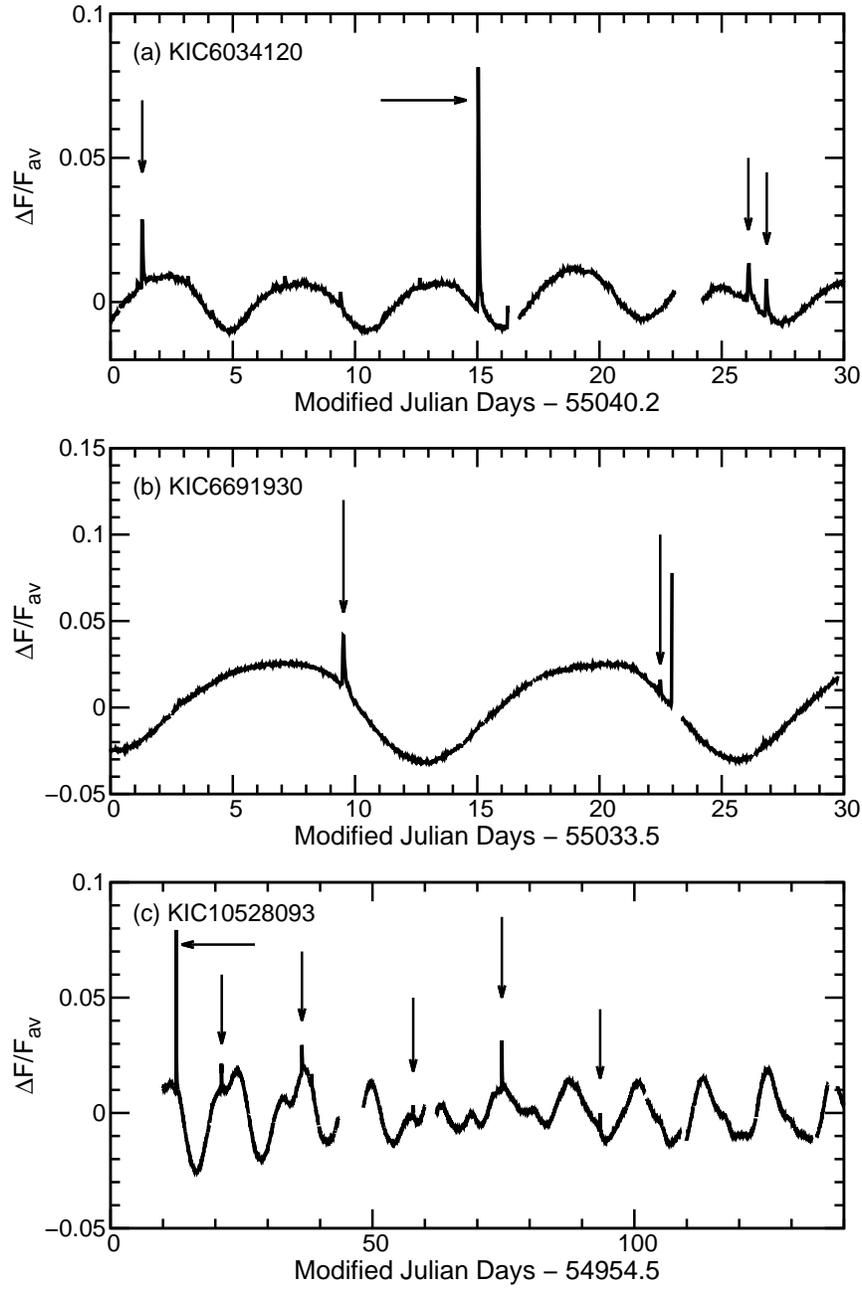}
\caption{Light curve of typical superflares on solar-type stars (\KICa \ \& \KICb, Quarter 2; \KICc, Quarter 0-2). 
The vertical axis is the brightness variations relative to the average brightness. 
The typical photometric error is about 0.02 \%. 
Arrows indicate superflares reported in \citet{maehara2012}.}
\label{lc}
\end{center}
\end{figure}

\begin{figure}[htpb]
\begin{center}
\epsscale{0.6}
\plotone{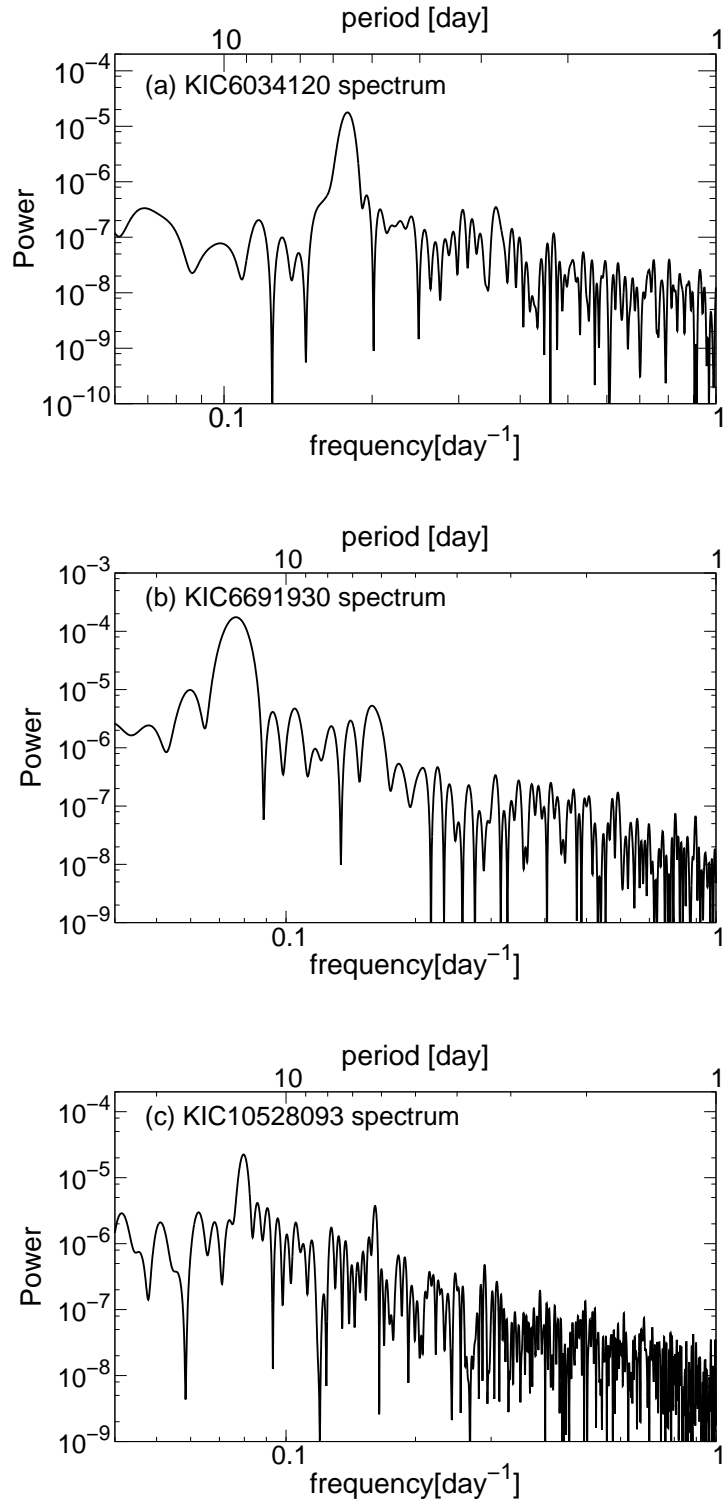}
\caption{Power spectra of the light curves shown in Figure \ref{lc}. 
The periods of these three stars are shown in Table \ref{table1}.}
\label{period_ana}
\end{center}
\end{figure}

\begin{figure}[htbp]
\begin{center}
\epsscale{0.6}
\plotone{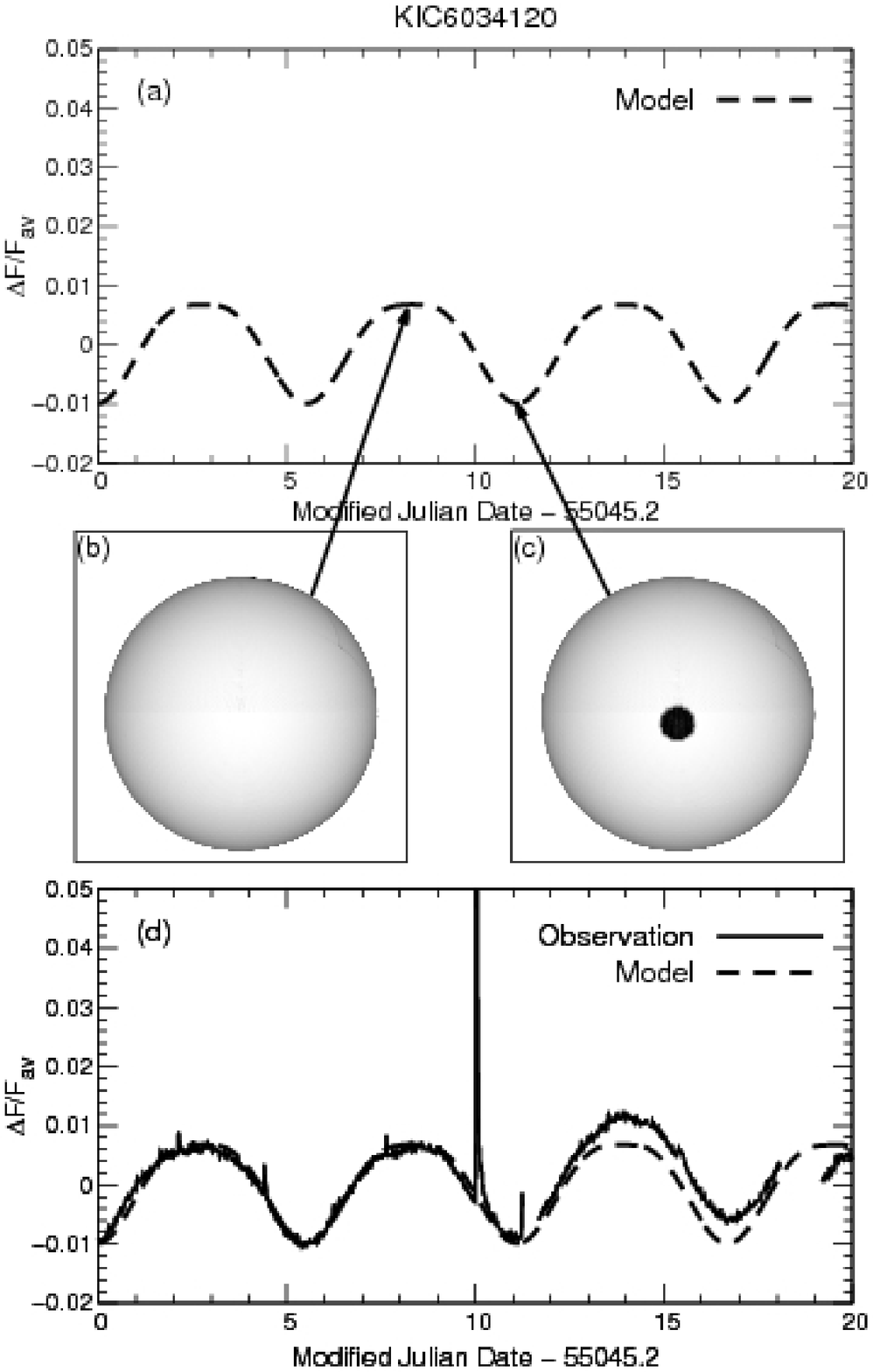}
\caption{(a): Model light curve for \KICa \ (Figure \ref{lc} (a)). 
The model parameters are given in Table \ref{table2}.  
(b) \& (c): Model pictures of the visible area of the photosphere with a starspot.  
(d): Observed light curve (solid line; the same as in Figure \ref{lc} (a)) and the model one (dashed line) for \KICa.}
\label{m\KICa}
\end{center}
\end{figure}

\begin{figure}[htbp]
\begin{center}
\epsscale{0.6}
\plotone{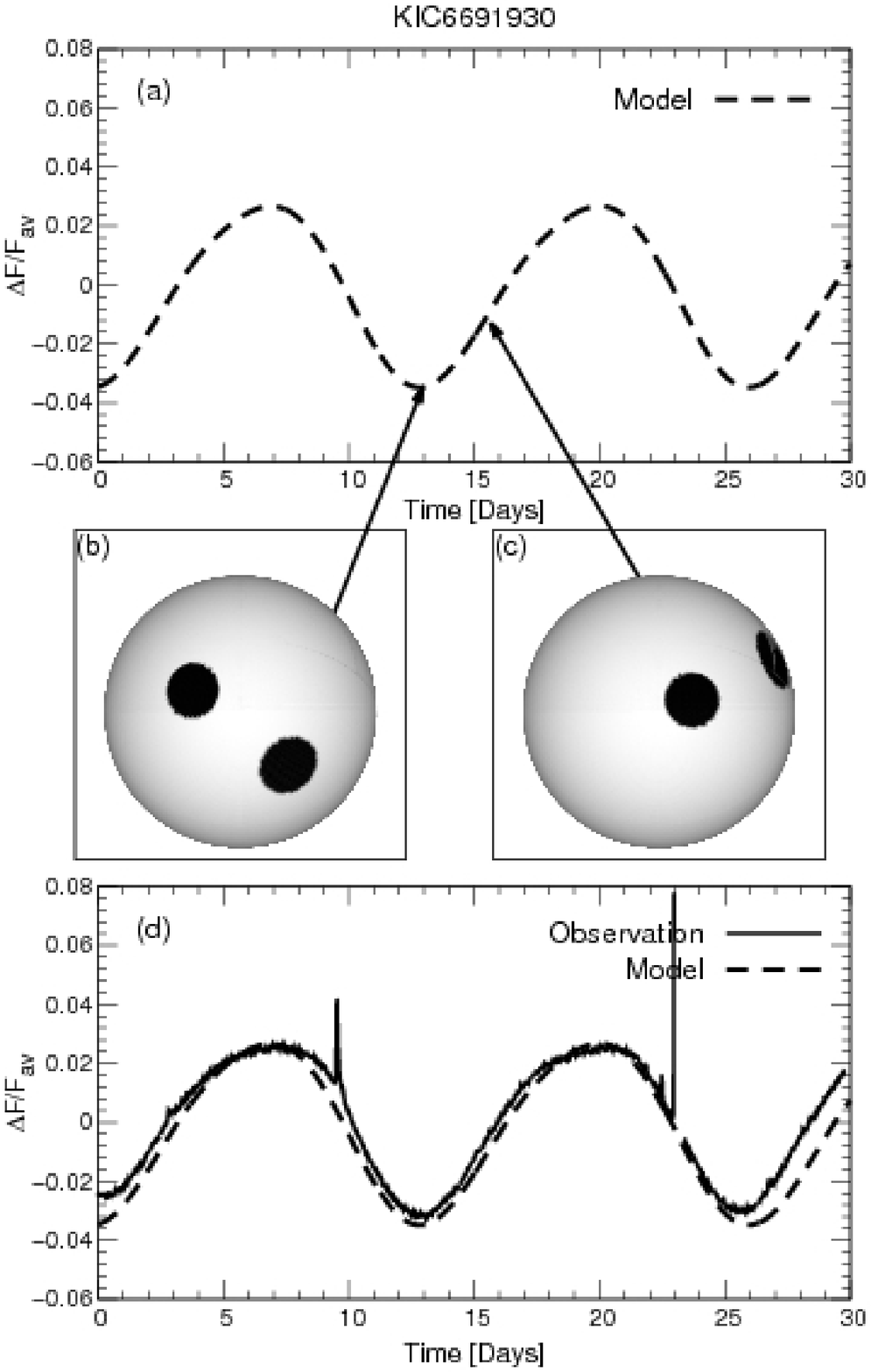}
\caption{(a) : Model light curve for \KICb. 
The model parameters are given in Table \ref{mo_para}. 
(b) \& (c): Model pictures of the visible area of the photosphere with two starspots. 
(d): Observed light (solid line; the same as in Figure \ref{lc} (b)) and the model one (dashed line) for \KICb. }
\label{m\KICb}
\end{center}
\end{figure}

\begin{figure}[htbp]
\begin{center}
\epsscale{0.6}
\plotone{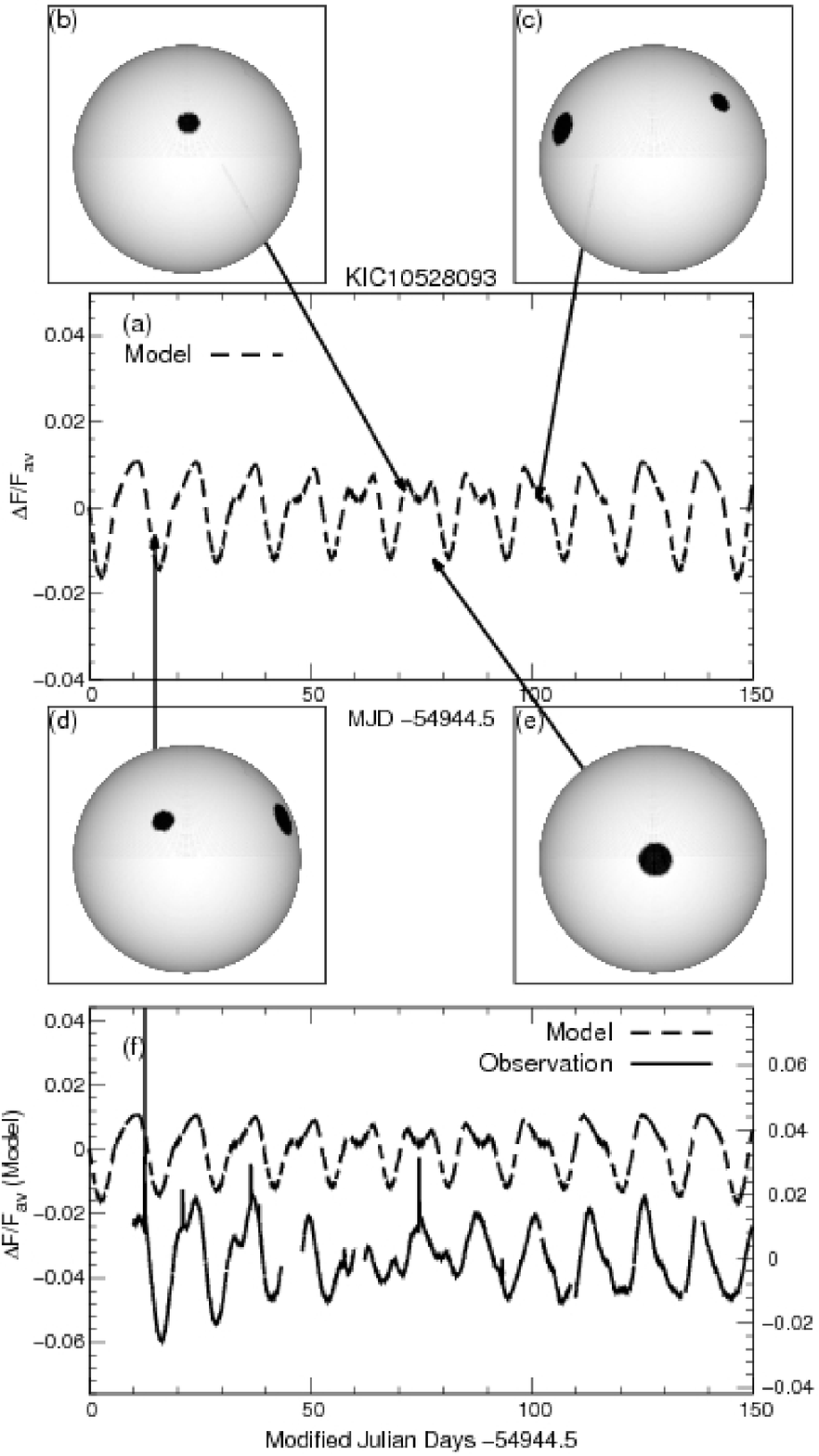}
\caption{(a) Model light curve for \KICc. 
The model parameters are given in Table \ref{mo_para}. 
(b), (c), (d) \& (e): Model pictures of the visible area of the photosphere with starspots.  
(f): Observed light curve (solid line; the same as in Figure \ref{lc} (c)) and the model one (dashed line) for \KICc.}
\label{m\KICc}
\end{center}
\end{figure}

\begin{figure}[htbp]
\begin{center}
\epsscale{0.7}
\plotone{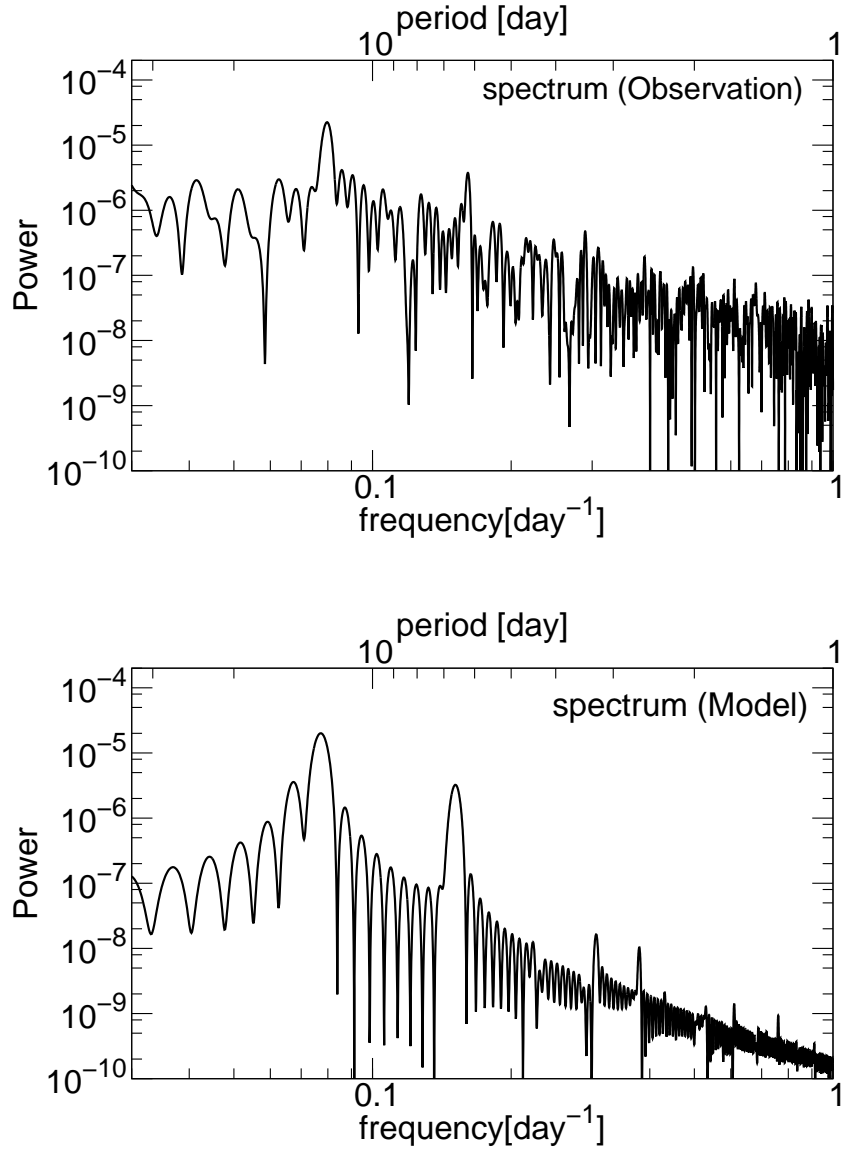}
\caption{Power spectra of the period 
analysis for the observed light curve and for the model one of \KICc \ shown in Figure \ref{m\KICc}(f).}
\label{fr-lm\KICc}
\end{center}
\end{figure}

\noindent
We performed simple calculations to show that brightness variations of superflare-generating stars can be explained by their rotation and the presence of large starspots.
There are many former researches which calculate the brightness variation by assuming the existence of 
dark spots on the surface of the star (e.g., \citealt{budding1977}; \citealt{dorren1987}; \citealt{eker1994}). 
Spot modeling has much uncertainty 
and it is almost impossible to get unique solutions from the spot modeling 
(e.g., \citet{eker1996}; \citet{kovari1997}; \citet{walkowicz2013}).
There are many parameters such as inclination angle and spot latitude which affect the shape of the lightcurve. 
In addition, the Kepler spacecraft provides only single color photometry data, and this also generates degeneracy 
between the spot size and the spot contrast.  
Our main purpose is to demonstrate that the brightness 
variations include the information of the overall areal coverage by starspots and the rotation periods of superflare stars. 
\\ \\
\ \ Our method of the spot modeling is as follows. 
We regard the total luminosity of the photosphere, and the starspot area as a good representation of 
the flux ($F$) in our modeling. 
The shape of the model star is assumed to be a sphere, 
and some circular spots are placed on the photosphere. 
It should be noted that we do not exclude the possibility 
that starspots of superflare stars are collections of smaller spots. 
The surface of the star is divided into 180 pieces ($j$) in the latitudinal angle and into
360 pieces ($k$) in the longitudinal angle. 
The contributions to the total brightness are summed up from each piece ($f_{j,k}$) as follows; 
\begin{equation} \label{method1}
F\propto\sum_{j,k}f_{j,k} \ ,
\end{equation}
and
\begin{equation} \label{method2}
f_{j,k}= \left \{
\begin{array}{l}
f_{j,k}^{\rm spot} (\rm{in \ the \ spot \ area}) \\
f_{j,k}^{\rm phot} (\rm{out \ of \ the \ spot \ area}) \ .
\end{array}
\right. 
\end{equation}
The temperature of the photosphere ($T_{\rm phot}$) and that of the spot area ($T_{\rm spot}$) are here 
considered to be the same over the solar photosphere (6,000K) 
and the sunspot area (4,000K, cf. \citealt{berdyugina2005}), respectively. 
The luminosity of each piece which is in the starspot area is estimated by 
\begin{equation} \label{method3}
f_{j,k}^{\rm spot}= \left \{
\begin{array}{l}
\sigma_{\rm SB} T_{\rm spot}^{4}S_{\rm pix} \ \vector{n_{j,k}}\cdot\vector{a_{j,k}} \ (\vector{n_{j,k}}\cdot\vector{a_{j,k}}\geq 0) \\
0 \ (\vector{n_{j,k}}\cdot\vector{a_{j,k}}< 0) \ ,
\end{array}
\right.
\end{equation}
where $S_{\rm pix}$ and $\vector{n_{j,k}}$ is a size and a normal vector of each pixel, respectively, and 
$\vector{a_{j,k}}$ is a unit vector which is antiparallel to the line of sight. 
$\sigma_{\rm SB}$ is the Stefan-Boltzmann constant. 
The luminosity of each piece which is out of the starspot area is estimated by 
\begin{equation} \label{method4}
f_{j,k}^{\rm phot}= \left \{
\begin{array}{l}
\sigma_{\rm SB} T_{\rm phot}^{4}S_{\rm pix} \ \vector{n_{j,k}}\cdot\vector{a_{j,k}} \ (\vector{n_{j,k}}\cdot\vector{a_{j,k}}\geq 0) \\
0 \ (\vector{n_{j,k}}\cdot\vector{a_{j,k}}< 0) \ .
\end{array}
\right.
\end{equation}
We also take into the effect of limb darkening.
Applying the equation (1) in \citet{sing2010}, 
this effect is calculated by 
\begin{equation} \label{method5}
\frac{f_{j,k}(\mu)}{f_{j,k}^{\rm NLD}}=1-u(1-\cos{\mu}) \ ,
\end{equation}
where $f_{j,k}^{\rm NLD}$ is the luminosity of each piece without the limb-darkening effect, 
$f_{j,k}(\mu)$ is that with the limb-darkening effect, 
$\mu=\cos{\Theta }$ and $\Theta$ is the viewing angle.
We here assume $u=0.6$ on the basis of Figure 3 in \citet{claret2004}. 
Because of the rotation, the visibility of the starspots changes, and the brightness then varies. 
We calculate the brightness by changing the angle of rotation by 1 degree. 
Model light curves are drawn with arbitrarily changing the spot position, spot radius, and inclination angle $i$, 
which is the angle between the line of sight and the rotation axis of the star. 
We selected cases which show rough agreement between the model and the observation  by eye.
\\ 
\\
\ \ Figures \ref{m\KICa}, \ref{m\KICb}, and  \ref{m\KICc} show results of the spot model 
calculation for the brightness variations of typical solar-type stars 
having superflares in Figure \ref{lc} (\KICa, \ \KICb, and \KICc). 
The stellar parameters of these stars are shown in Table \ref{st_para}. 
The period of the brightness variations of these stars is about 5$\sim$15 days, 
and the amplitude is from 0.1\% to a few \%. 
\\ \\
\ \ As shown in Figure \ref{m\KICa}, the simple sine-like light curve of \KICa \ 
can be reproduced by the model with one spot. 
Figure \ref{m\KICa} (a) shows the model light curve for \KICa. 
The best set of model parameters are listed in Table \ref{mo_para}. 
Snapshots of the modeled star are displayed in Figure \ref{m\KICa} (b) and (c), 
which shows the star has a large spot compared to the Sun. 
Figure \ref{m\KICa} (d) shows the comparison of observed light curve and the model light curve. 
\\ \\
\ \ The shape of the light curve of \KICb \ has one peak and one shoulder as shown in Figure \ref{lc} (b). 
The features can be reproduced by the two spot model as shown in Figure \ref{m\KICb}. 
The longitude difference between two spots can be estimated by measuring 
the rotational phase of the shoulder in the light curve. 
Figure \ref{m\KICb} (a) shows the model light curve for \KICb. 
The model parameters are listed in Table \ref{mo_para}. 
Figure \ref{m\KICb} (b), (c), and (d) are drawn for \KICb \ 
in the same manner of Figure \ref{m\KICa} (b), (c), and (d) for \KICa. 
\\ \\
\ \ The shape of the light curve of \KICc \ also has a shoulder as shown in Figure \ref{lc} (c). 
However, the long-term amplitude variation cannot be reproduced 
by a simple model with two spots in solid body rotation since the phase of the shoulder changes as 
time progresses.
\\ 
\\
\ \ These variations can be explained by the differential
rotation (e.g., \citealt{frasca2011}; \citealt{frohlich2012}). 
Here, we consider the two-spot model with the differential rotation. 
It should be noted that what this model shows is only that 
the differential rotation in this model is of approximately the correct amount. 
This is because a qualitatively similar fit would be produced by any models, 
with almost any spot parameters, which get the amount of the differential rotation correct 
and the rough longitudinal separation correct.
Parameters of the differential rotation (the dependence of the rotation speed on the latitude) we use here 
is based on the value for the Sun \citep{snodgrass1990}. 
The relation  between the angular velocity of the rotation ($\omega$) 
and the latitude ($\varphi$) is represented by the equation (\ref{dr-para}).
\begin{eqnarray} \label{dr-para}
\omega =A+B\sin^{2}\varphi+C\sin^{4}\varphi \ . 
\end{eqnarray}  
For the Sun, the values of $A$, $B$ and $C$ are as follows; 
$A = 14.71 \ \rm{deg \ day}^{-1}$, $B = -2.39 \ \rm{deg \ day}^{-1}$, 
and $C = -1.78 \ \rm{deg \ day}^{-1}$ \citep{snodgrass1990}. 
We here ignore the ``$C\sin^{4}\varphi$" term since this term is negligibly small in our model, 
and then we convert values $A$ and $B$ to 
ones which corresponds to the rotation period of \KICc. 
For \KICc, the value of $A$ and $B$ are 
as follows; $A = 31.1 \ \rm{deg \ day}^{-1}$, and $B = -5.05 \ \rm{deg \ day}^{-1}$. 
Figure \ref{m\KICc} (a) shows the model light curve for \KICc.
The model parameters are listed in Table \ref{mo_para}.
Snapshots of the modeled star are shown in Figure \ref{m\KICc} (b), (c), (d), and (e). 
We compare the observed light curve with modeled light curve in Figure \ref{m\KICc} (f).  
Figure \ref{fr-lm\KICc} shows power spectra of the observed light curve 
and the model light curve in Figure \ref{m\KICc} (f). 
We can see that the feature of the power spectrum, that is, the dominant and overtone peaks, 
and the slope to the higher frequencies, is also roughly  reproduced 
with our spot model. 
\\ \\ 
\ \ Consequently, most of the brightness variations with the period of a few days to a few tens of days 
and the amplitude of 0.1 $\sim$ 10\% 
are expected to be able to be explained by assuming that the star has fairly large starspots. 
\\
\section{Ensemble Properties of Superflares: Dependence of Flare Energy and Flare Frequency on Rotation Period}\label{sec:period-ana-plot}
\begin{figure}[htpb]
\begin{center}
\epsscale{0.45}
\plotone{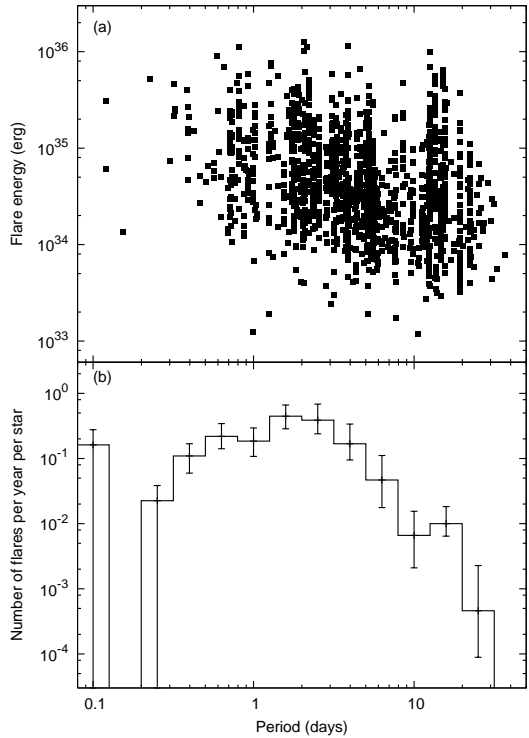}

\caption{{\scriptsize (a) Scatter plot of the flare-energy vs the brightness variation period. 
The period of the brightness variation in this figure was estimated by using the Kepler data of Quarter 1-6. 
We do not use Quarter 0 data here since the data period is shorter ($\sim$11 days) than other Quarters and sometimes 
long term variations ($\sim$30 days) are removed. 
An apparent negative correlation between the variation period 
and the lower limit of the flare energy results from the detection limit 
of our flare-search method. 
(b) Distribution of the occurrence of flares in each period bin as a function of the brightness variation period. 
The vertical axis indicates the number of flares with energy $\geq 5\times10^{34}$ erg per star per year. 
The error bars represent the 1$\sigma$ uncertainty estimated from 
the square root of the event number of flares in each bin. 
The frequency distribution of superflares saturates 
for periods shorter than a few days. 
A similar saturation is observed for the relations between the coronal X-ray activity 
and the rotation period \citep{randich2000}.
Note that these figures are a bit different from those of Figure 3 of \citet{maehara2012}; 
the number of flares per year per stars for stars with period between 20 and 30 days is about 30 times smaller than that of \citet{maehara2012}. Main reason of this difference is that the number of solar-type stars 
in longer period bins is larger than that in \citet{maehara2012}. 
This is because the light curves produced by the improved pipeline (PDC-MAP; e.g., \citet{stumpe2012} were 
used for the period analysis and the long-period brightness variations can be more easily detected 
in the improved light curves. 
The number of superflare stars did not change so much because these stars tend to have large starspots 
and it was easy to detect the long term brightness variations even in \citet{maehara2012}, which did not use PDC-MAP pipeline. (See \citet{shibayama2013} for more details.)}}
\label{period_plot}
\end{center}
\end{figure}

\noindent
Applying the results shown in Section \ref{sec:model-ana}, we assume that many of the brightness variations of solar-type stars with superflares can be explained by their rotation and the presence of large starspots.
Only three particular lightcurves were shown to be roughly reproduced 
by our uncomplicated model calculations in the above section. 
This does not prove that such spot models can be applied to the lightcurves of all the superflare stars. 
\\ \\ 
\ \ Figure \ref{period_plot} shows the relations between the rotation period and the features of superflares. 
Although the similar relations  have been already discussed in \citet{maehara2012}, we here add 
data of superflares we newly found in \citet{shibayama2013} from Kepler quarter 3-6 data 
and improve the statistical precision by increasing the numbers of data. 
The rotation periods are determined by using the way described near the end of section \ref{sec:Kepler-data}, 
although this way includes some uncertainty (e.g., \citealt{walkowicz2013}). 
Figure \ref{period_plot} (a) indicates that the most energetic flare observed 
in a given rotation period bin does not have a correlation with the period of stellar rotation. 
If the superflare energy can be explained by the magnetic energy stored near the starspots 
(we discuss this in Section \ref{sec:discussion}), 
this result suggests that the maximum 
magnetic energy stored near the spot does not have a strong dependence
on the rotation period. 
Figure \ref{period_plot} (b) shows that the average flare frequency in a given period bin tends
to decrease as the period increases to periods longer than a few days. 
The value of the frequency is the average of all superflare stars in the same period bin. 
Some of superflare stars (e.g., examples in Figure \ref{lc}) are then able to have higher frequencies of superflares.
The frequency of superflares on rapidly rotating stars 
is higher than slowly rotating stars. 
It is known that the rotation period correlates with the chromospheric activity 
and the more rapidly rotating stars 
have higher magnetic activity (e.g., \citealt{noyes1984} ; \citealt{pallavicini1981}). 
These imply that rapidly rotating stars with high magnetic activity 
can generate more frequent superflares. 
\\ \\
\section{Discussion on starspot coverage and Energy of superflares}\label{sec:discussion}
\begin{figure}[htbp]
\begin{center}
\epsscale{0.7}
\plotone{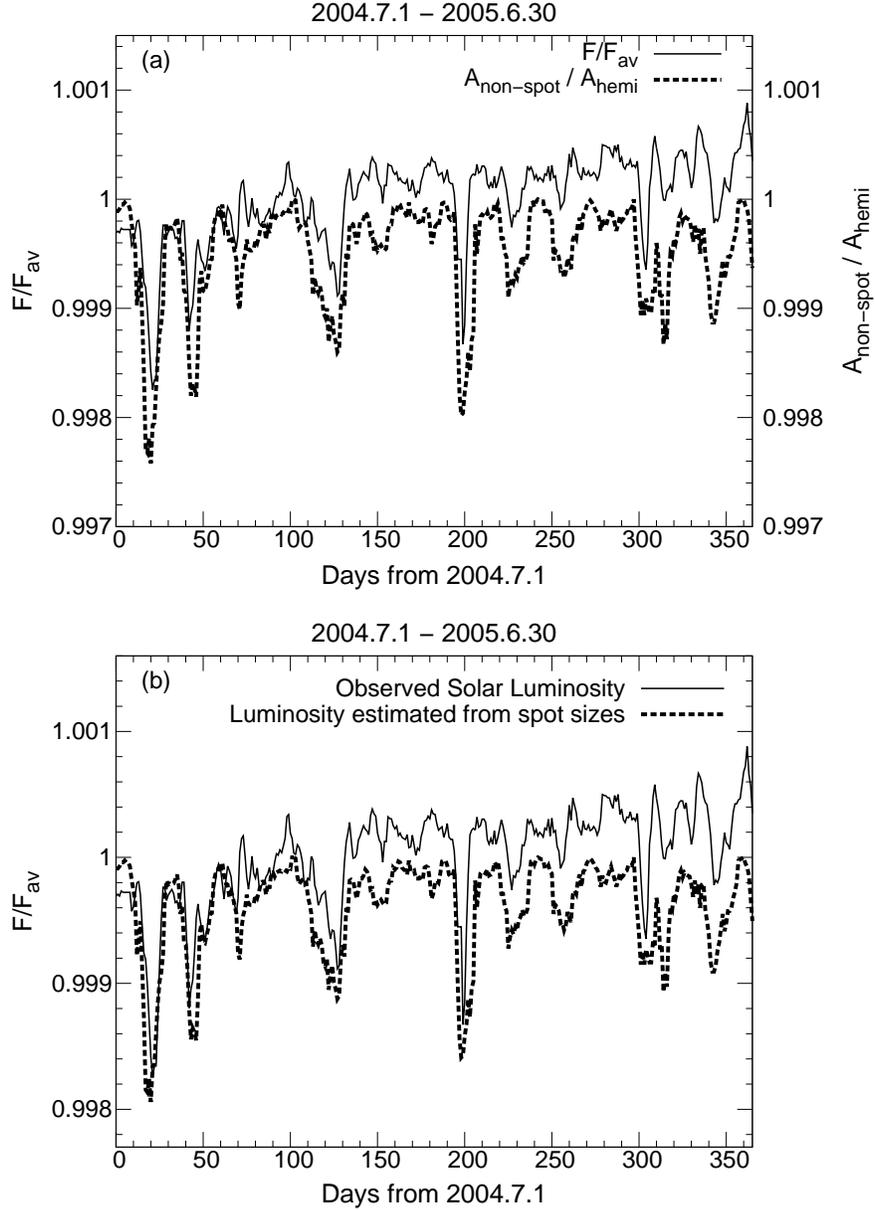}
\caption{(a): Normalized irradiance of the Sun and area not coverd by sunspots. 
The data period is 2004.7.1 - 2005.6.30. Solid lines are normalized visible (4,500-8,000{\AA}) solar irradiance 
from the solar spectral irradiance data of the SORCE Satellite. 
We use the daily sunspots area data prepared by the U.S. Dept. of Commerce, NOAA, Space Weather Prediction Center. 
(b): Normalized irradiance of the Sun and the luminosity estimated 
from the spot coverages assuming that the temperature of sunspots is 4,000K 
and that of the solar photosphere is 6,000K. The data period is the same as (a). }
\label{sun_lcm}
\end{center}
\end{figure}

\begin{figure}[htbp]
\begin{center}
\includegraphics[angle=-90,scale=.60]{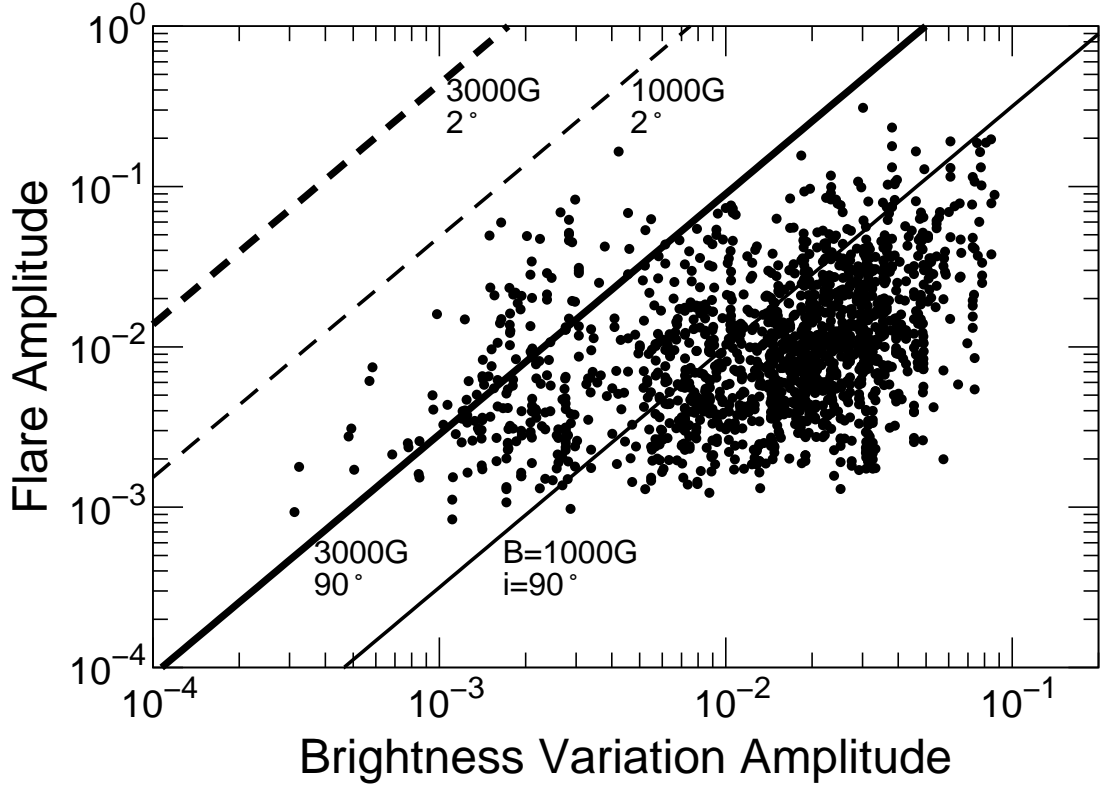}
\caption{Scatter plot of superflare amplitude as a function of the amplitude of the brightness variation. 
The data of superflares are taken from quarter 1-6 data 
and are reported in \citet{shibayama2013}.
We here defined the amplitude as 
the normalized brightness range, in which the lower 99 percent of the distribution of
brightness difference from the average, except for the flares, are included. 
Thick and thin solid lines correspond to
the analytic relation between the stellar brightness variation amplitude (corresponding to
the spot area) and flare amplitude (correspond to the flare energy) obtained 
from equation (\ref{energy7}) for $B$=3,000G and
1,000G. The thick and thin dashed lines correspond to 
the same relation in case of nearly pole-on (i=2.0 deg) for $B$=3,000G and 1,000G. 
These lines are considered to give an upper limit 
for the flare amplitude.}    
\label{vamp_lamp} 
\end{center} 
\end{figure}

\begin{figure}[htbp]
\begin{center}
\epsscale{0.55}
\plotone{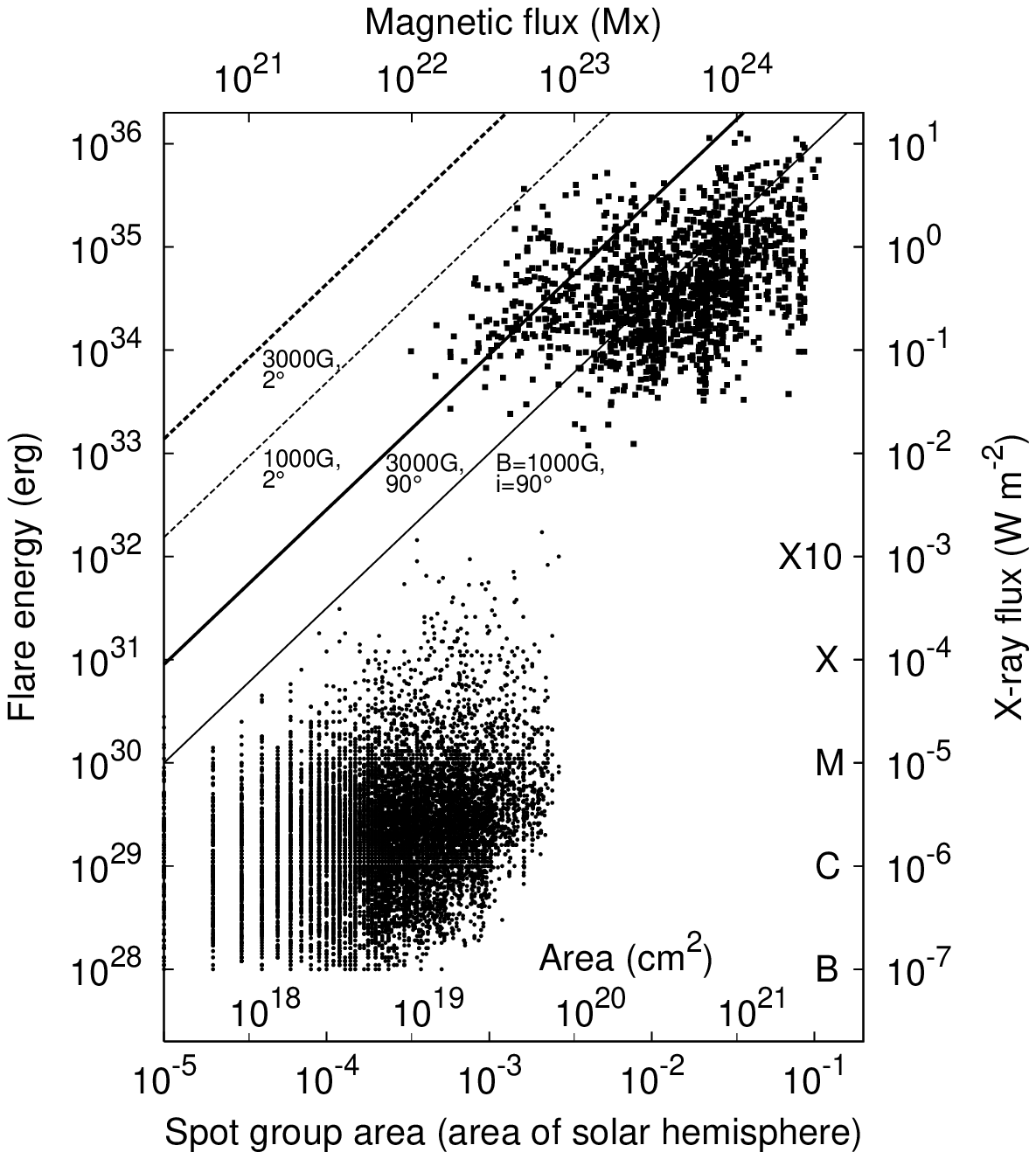}
\caption{Flare energy vs spot coverage of 
superflares on solar-type stars (filled squares; \citealt{shibayama2013}) 
and solar flares (filled circles; \citealt{sammis2000}, \citealt{ishii2012}). 
The data of solar flares are taken from ftp://ftp.ngdc.noaa.gov/STP, 
and consists of data in 1989-1997 \citep{sammis2000} and those in 1996-2006 \citep{ishii2012}. 
Thick and thin solid lines corresponds to
the analytic relation between the flare energy obtained from equation (\ref{energy8}) for $B$=3,000G and
1,000G. The thick and thin dashed lines correspond to 
the same relation in case of nearly pole-on (i=2.0 deg) for $B$=3,000G and 1,000G. 
These lines are considered to give an upper limit 
for the flare energy (i.e., possible maximum magnetic energy which can be stored near sunspots).
Note that the superflare on solar-type stars is observed only with visible light and 
the total energy is estimated from such visible light data. 
Hence the X-ray intensity in the right hand vertical axis is not based on actual observations. 
The energy of solar flares is based on the assumption that 
the energy of X10-class flare is $10^{32}$ erg, X-class $10^{31}$ erg, 
M-class $10^{30}$ erg, and C-class $10^{29}$ erg, considering 
previous observational estimate of energies of typical solar flares (e.g., \citealt{benz2010}). 
The values on the horizontal axis at the top show the total magnetic flux of spot 
corresponding to the area on the horizontal axis at the bottom when $B$=1,000G.}   
\label{amp_sf_sun}    
 \end{center} 
\end{figure}

\noindent As described  in Section \ref{sec:intro}, it has been said that the brightness variation 
of many solar-type stars observed by Kepler is due to the rotation of the star with starspots. 
This effect of rotation is also seen in the Sun. Figure \ref{sun_lcm} represents that there are 
also the same relations in the Sun; changes of the visible area of the sunspot causes the brightness variation. 
However, faculae also can affect the brightness variations of the Sun (e.g., \citealt{lanza2003}). The brightness variations of superflare-generating stars are also probably somewhat affected by faculae, 
though \citet{lockwood2007} indicates that the photometric behaviors of (young) active stars are dominated by starspots. In this paper, we do not include the faculae effects when analyzing the Kepler data, as described in \citet{walkowicz2013}.
\\ 
\\
\ \ \citet{maehara2012} showed that superflare-generating stars have 
the brightness variation with the period of a few days to several tens of days, and 
that many of such variations are likely due to the rotation, though the possibility that 
the star is a binary is not completely excluded. 
It is well known that more than half of general solar-type stars are binary (e.g., \citealt{duquennoy1991}). 
However photometric observations by Kepler mission cannot detect spectroscopic binaries. 
We cannot exclude the possibility that superflares are taking place on the spectroscopic binary companions. 
Because of this, the period data shown in Figure 7 are not necessarily due to rotation, 
and in particular the target with a 0.1-day periodicity is probably a binary. 
Spectroscopic observation is necessary, which will be our future project (e.g., \citealt{snotsu2013}).
\\ \\
\ \ Here we assume the brightness varitations are due to the rotation. 
Our model analyses for some typical brightness variations  
suggest us that, if we take into account the effects of some parameters, such as, the inclination angle, and the spot latitude, 
the brightness variations whose amplitude is about 0.1$\sim$10\% can be 
well explained by the rotation of the star with fairly large starspots. 
\\ 
\\
\ \ In the following, we discuss the relation 
between the starspot coverage and the energy of superflares. 
Flares are the release of the stellar magnetic energy (e.g., \citealt{shibata2011}). 
It is known that there are positive correlations between the sunspot coverage 
and the energy of the largest solar flares \citep{sammis2000}.
It is, therefore, important to consider whether the same correlations 
can be found and the observed maximum of superflare energy 
can be explained by the magnetic energy 
of large starspots which many of the superflare stars are expected to have.
Although we have already simply discussed the relation between the amplitude of 
superflares and that of the brightness variations 
in  Supplementary Information Section of  \citet{maehara2012}, 
we here advance this former analysis and discuss whether 
the magnetic energy of starspots can well explain 
the energy of superflares by applying the results in this paper 
that starspot coverages can be roughly estimated from the brightness variation.
\\ \\
\ \ First, we discuss expected relations if the energy 
sources of superflares are the magnetic energy stored around the starspots ($E_{\rm flare}\leq E_{\rm mag}$). 
The total energy released by the flare must be smaller than (or equal to) the magnetic 
energy stored around the starspots. 
The order of the stored magnetic energy ($E_{\rm mag}$) can be roughly estimated by
\begin{equation} \label{energy1}
E_{\rm mag}\approx \frac{B^{2}l^{3}}{8\pi} \ ,
\end{equation}
where $B$ and $l$ correspond to the magnetic field strength and the size of the starspot region. 
It should be noted that we cannot exclude the possibility that starspots of superflare stars are collections of smaller spots. 
\\ 
\\
\ \ 
We roughly assume that there is a linear correlation between the brightness variation and the spot coverage 
when we estimate the magnetic energy stored 
around the starspots from the amplitude of the brightness variation. 
The total amplitude of the brightness variation due to the rotation normalized 
by the average brightness ($\Delta F_{\rm rot}$) can be expressed by 
\begin{equation} \label{energy2}
\Delta F_{\rm rot}\approx\biggl[1-\Bigl(\frac{T_{\rm spot}}{T_{\rm phot}}\Bigl)^{4}\biggl]\frac{A_{\rm spot}}{A_{\rm phot}} \ ,
\end{equation}
where $A_{\rm spot}$ is the total area of starspots, and $A_{\rm phot}$ is the total visible area of the stellar surface. 
Since the Kepler photometer covers a wide spectral rage (4,000 $\sim$ 8,500\AA), observed brightness changes 
would be nearly approximated by the bolometric brightness change obtained by the equation (\ref{energy2}). 
\\ 
\\
\ \ If we assume that $l^{3} \approx A_{\rm spot}^{3/2}$, the equation (\ref{energy1}) can be transformed to 
\begin{equation} \label{energy3}
E_{\rm mag}\approx \frac{B^{2}}{8\pi}A_{\rm spot}^{3/2}\approx \frac{B^{2}}{8\pi}\biggl(\frac{A_{\rm phot}\Delta F_{\rm rot}}{1-(T_{\rm spot}/T_{\rm phot})^{4}}\biggl)^{3/2}\geq E_{\rm flare} \ ,
\end{equation}
where $E_{\rm flare}$ is the total energy released by the flare. 
The amplitude of the flare normalized by the average brightness $(\Delta F_{\rm flare})$ can be estimated by 
\begin{equation} \label{energy4}
\Delta F_{\rm flare} \approx \frac{E_{\rm flare}}{L_{\rm star}\tau} \ ,
\end{equation}
where $L_{\rm star}$ is the luminosity of the star, and $\tau$ is the e-folding time of the flare. 
Then the relation  between the flare amplitude and the brightness variation amplitude can be written as
\begin{equation} \label{energy5}
\Delta F_{\rm flare} \leq \frac{B^{2}}{8\pi}\frac{A_{\rm spot}^{3/2}}{L_{\rm star}\tau}\approx \frac{B^{2}}{8\pi}\frac{1}{L_{\rm star}\tau}\biggl(\frac{A_{\rm phot}\Delta F_{\rm rot}}{1-(T_{\rm spot}/T_{\rm phot})^{4}}\biggl)^{3/2}\propto A_{\rm spot}^{3/2} \propto (\Delta F_{\rm rot})^{3/2} \ .
\end{equation}
This indicates that the upper limit of 
the flare amplitude is proportional to 1.5 power of the starspots area if we assume $B=constant$. 
We discuss in the following whether this assumption is right. 
\\ \\
\ \ The discussion described around equations (\ref{energy1})$\sim$(\ref{energy5}) does not take into account 
the effect of the inclination angle $i$ and spot latitude. 
In stars with lower inclination angle and higher spot latitude, 
the amplitude of brightness variation is smaller. 
If we consider inclination effects and assume the spot is distributed around the equator, equation (\ref{energy5}) 
can be replaced by equation (\ref{energy6}). 
\begin{equation} \label{energy6}
\Delta F_{\rm flare} \leq \frac{B^{2}}{8\pi}\frac{(A_{\rm spot}/\sin{i})^{3/2}}{L_{\rm star}\tau}\approx \frac{B^{2}}{8\pi}\frac{1}{L_{\rm star}\tau}\biggl(\frac{A_{\rm phot}\Delta F_{\rm rot}}{1-(T_{\rm spot}/T_{\rm phot})^{4}}\frac{1}{\sin{i}}\biggl)^{3/2} \ .
\end{equation}
If we adopt $L_{\rm star}=10^{33}$erg s$^{-1}$, $\tau=10^{4}$sec, $T_{\rm spot}=4,000$K and $T_{\rm phot}=6,000$K 
for superflares on solar-type stars, then the flare amplitude can be roughly estimated by 
\begin{equation} \label{energy7}
\Delta F_{\rm flare} \propto 10\biggl(\frac{B}{1000\rm{G}}\biggl)^{2}\biggl(\Delta F_{\rm rot}\frac{1}{\sin{i}}\biggl)^{3/2} \ .
\end{equation}
\\ \\
\ \ Figure \ref{vamp_lamp} is the scatter plot of the superflare amplitude 
as a function of the amplitude of the brightness variation. 
The data of superflares are taken from quarter 1-6 data 
and are reported in \citet{shibayama2013}. 
The lines in this figure represent the analytic relation 
between the flare amplitude and the amplitude of the brightness variation 
obtained from equation (\ref{energy7}) for $i$=90 degree and $i$=2 degree (nearly pole-on) case each. 
Taking into account that magnetic energy density of solar-type 
stars is about $1,000\sim4,000$G \citep{solanki2003},  
we show two lines ($B$=1,000G \ \& \ 3,000G) each. 
These lines give an upper limit for the flare amplitude based on the above discussion.
Many of the flares
have smaller amplitudes than that expected by the equation (\ref{energy7}). 
Moreover all flares have smaller flare
amplitudes than that expected in the nearly pole-on case (dashed-line; $i$=2 degree). 
This supports that the energy released 
by these flares can be explained by the magnetic energy stored
near the starspots. 
\\ \\
\ \ For the solar flares, there is a similar relation as this \citep{sammis2000}. 
Figure \ref{amp_sf_sun} shows the empirical relation between the spot group area 
and X-ray intensity of solar flares \citep{sammis2000}, 
and the relation between the spot group area (estimated from the brightness variation amplitude) 
and the superflare energy. 
It should be noted that while there is a large range in observed flare energies for a given sunspot area, 
we here in particular refer to the upper envelope of the observations shown in Figure \ref{amp_sf_sun}, where the largest flare energies are associated with larger spots. 
The bolometric luminosity and the total bolometric energy of superflares were estimated from 
the stellar radius, the effective temperature in the Kepler Input Catalog \citep{brown2011}, 
observed amplitude and duration of flares by 
assuming that the spectrum of white light flares 
can be described by the blackbody radiation with the effective 
temperature of $\sim$10,000K (\citealt{kretzschmar2011}). 
Using concrete numerical values, equation (\ref{energy3}) is transformed as  
\begin{eqnarray} \label{energy8}
E_{\rm mag} & \approx & \frac{B^{2}}{8\pi}A_{\rm spot}^{3/2} \nonumber\\
& \approx & 10^{33}[erg] \biggl(\frac{B}{10^{3}\rm{G}}\biggl)^{2}\biggl(\frac{A_{\rm spot}}{3\times 10^{19}\rm{cm}^{2}}\biggl)^{3/2} \nonumber\\
&\approx & 10^{33}[erg] \biggl(\frac{B}{10^{3}\rm{G}}\biggl)^{2}\biggl(\frac{A_{\rm spot}/(2\pi R_{\sun}^{2})}{0.001}\biggl)^{3/2} \nonumber\\
& \geq & E_{\rm flare}  \ ,
\end{eqnarray}
where $R_{\sun}$ is the radius of the Sun. 
The relation expressed by equation (\ref{energy8}) can be seen in Figure \ref{amp_sf_sun}. 
All of the solar flare data, and over a half of the superflare data 
are located below the same line of $B$=1,000G, and $i$=90deg. 
\\ 
\\
\ \ Some data points of superflares are located above the line of equation (\ref{energy8}). 
This is explained by the two reasons discussed in the following. 
First, the inclination and latitude of the spots govern their projected area, 
and the size of the spot cannot be uniquely determined 
because of these effects and the fixed contrast with the photosphere. 
Lower inclinations cause a smaller projected spot area and/or the spot to be visible 
through most of the rotation period, and therefore lower brightness variations result in these cases.
\\ \\
\ \ Second, magnetic flux density ($B$) is not necessarily constant. 
It is a function of  the spot size. 
\citet{solanki2003} shows that the umbral continuum intensity of sunspots becomes weak as 
sunspots become large and that the intensity also becomes weak as the magnetic flux density gets strong. 
This fact means that as the spot becomes larger, the magnetic flux density becomes strong and 
the magnetic energy stored near the spot increases. 
As a result, the released flare energy tends to be bigger than that in the case of $B$=$constant$, and the 
straight lines in Figure \ref{amp_sf_sun} become steeper. 
\\ 
\\
\ \ In this paper, we showed that many of superflare-generating stars probably have large starspots, 
and the energy of superflares can be explained by the magnetic energy stored around such starspots. 
We also confirmed in Section \ref{sec:period-ana-plot} the stellar activity depends on the rotation period. 
Stars with relatively slower rotation rates can still produce flares that are as energetic 
as those of more rapidly rotating stars, although the average flare frequency is lower for more slowly rotating stars.
There are, however, many things to be left for consideration in the next step.
For example, it is extremely important to know lifetime of starspot regions or how activity level changes 
because it is related to knowing how the starspot evolves by the dynamo mechanism. \citet{shibata2013}  
indicated that the magnetic energy to generate superflares can be produced by the effects of 
differential rotation. We need to investigate in detail the effects of differential rotation on the brightness variation.
\\
\\
\ \ In addition, if superflares occur on the starspot area of a single star and the brightness
variation is caused by the rotation, the timings when superflares occur 
are likely to have a phase dependence, since superflares 
which we can observe need to occur when the starspots are on the visible side of the star. 
This phase dependence, however, is expected not to be necessarily true. 
For some pairs of spot distribution on the stellar surface and stellar inclination angle, 
some part of the area covered by the starspots is always on the visible side of the star. 
In these cases, it is probable that superflare occurs in any phase of the differential rotation. 
We need to study these effects in detail in the next step 
(cf.\citet{roettenbacher2013} discuss these effects for one K-type star.).
\\
\section{Summary}\label{sec:summary}
\noindent
In this paper, we investigated the brightness variations of 
superflare-generating stars in Kepler quarter 0-6 data by considering 
the signature of stellar rotation and the starspots. 
First, we performed simple spot model analyses for typical superflare-generating stars. 
Many of the brightness variations with the period of one day to a few tens of days and the  
amplitude of 0.1 $\sim$ 10\% can be explained by the rotation of the star with fairly large starspots, 
taking into account the effects of the inclination angle and the spot latitude. 
Next, using the result of the period analysis, 
we investigated the relations between the energy and frequency of superflares 
and the rotation period by assuming that 
the period of the brightness variation corresponds to the rotation period. 
Stars with relatively slower rotation rates can still produce flares that are as energetic 
as those of more rapidly rotating stars, although the average flare frequency is lower for more slowly rotating stars.
Last, we discussed the relations  between the starspot coverage and the energy of flares. 
If we assume that the brightness variations are due to the rotation, the energy of superflares can be explained 
by the magnetic energy stored near the large starspots. 
\\
\\ 
\acknowledgments
We are grateful to Prof. Kazuhiro Sekiguchi (NAOJ) for useful suggestions. 
We also thank the anonymous referee for helpful comments. 
Kepler was selected as the tenth Discovery mission. 
Funding for this mission is provided by the NASA Science Mission Directorate. 
The data presented in this paper were obtained from the Multimission Archive at STScI. 
This work was supported by the Grant-in-Aid from the Ministry of 
Education, Culture, Sports, Science and Technology of Japan (No. 25287039).

\clearpage

\begin{table}[htpb]
\begin{center}
\caption{Length of the observation period during each quarter and the number of
solar-type stars.}
\begin{tabular}{cccc}
\hline
Quarter&N\tablenotemark{a}&$\tau$[days]\tablenotemark{b}&Release\tablenotemark{c}\\
\hline
0&9511&11&14\tablenotemark{d}\\
1&75598&33&14\tablenotemark{d}\\
2&82811&90&14\tablenotemark{d}\\
3&82586&90&14\tablenotemark{d}\\
4&89188&90&14\tablenotemark{d}\\
5&86248&95&16\tablenotemark{e}\\
6&82052&90&16\tablenotemark{e}\\
\hline
\label{data_summary}
\end{tabular}
\tablenotetext{a}{\ Number of solar-type stars.} 
\tablenotetext{b}{\ Length of the observation period during each quarter.} 
\tablenotetext{c}{\ Release number of the data we use in this paper.}
\tablenotetext{d}{\ Kepler Data Release numbers 14 Notes \citep{KDRNotes14}}
\tablenotetext{e}{\ Kepler Data Release numbers 16 Notes \citep{KDRNotes16}}
\end{center}
\end{table}

\begin{table}[htpb]
\begin{center}
\caption{Stellar parameters for three solar-type superflare-generating stars of which we show the results of model analysis in Figure \ref{m\KICa}, \ref{m\KICb} \& \ref{m\KICc}.}
\label{table1}
\begin{tabular}{cccccccccc}
\hline
Kepler ID&$T_{\rm{eff}}$\tablenotemark{a}&log $g$\tablenotemark{a}&$R/R_{\sun}$\tablenotemark{a}&Period\tablenotemark{b}&Amplitude\tablenotemark{c}\\
\hline
6034120&5407&4.7 & 0.77 & 5.6 & 4.45$\times$10$^{-3}$\\
6691930&5348&4.5 & 0.95 &13.1& 1.29$\times$10$^{-2}$\\
10528043&5143&4.5 & 0.88 & 12.9 & 6.52$\times$10$^{-3}$\\
\hline
\label{st_para}
\end{tabular}
\\
\tablenotetext{a}{These data are taken from the Kepler Input Catalog \citep{brown2011}.} 
\tablenotetext{b}{The period of the largest amplitude component of the brightness variations in unit of day.} 
\tablenotetext{c}{The normalized amplitude of the dominant period.} 
\end{center}
\end{table}

\begin{table}[htbp]
\begin{center}
\caption{\rm{Best set of model parameters for Figure \ref{m\KICa}, \ref{m\KICb} \& \ref{m\KICc}.}}
\label{table2}
\begin{tabular}{cccccc}
\hline
Kepler ID & Inclination & spot name & R$_{\rm{spot}}$/R$_{\rm{phot}}$ 
& Initial Latitude & Initial Longitude Difference \\
\hline
6034120 & 45$^\circ$ & spot1 & 0.13 & 41$^\circ$N & -- \\
\hline
6691930 & 32$^\circ$ & spot1 & 0.23 & 30$^\circ$N & -- \\
-- & -- & spot2 & 0.21 & 60$^\circ$N & 70$^\circ$ \\
\hline
10528043 & 60$^\circ$ & spot1 & 0.15 & 0$^\circ$N & -- \\
-- & -- & spot2 & 0.10 & 37$^\circ$N & 120$^\circ$ \\
\hline
\label{mo_para}
\end{tabular}
\end{center}
\end{table}

\clearpage

\clearpage

\clearpage

\clearpage

\clearpage
\clearpage

\clearpage

\clearpage

\clearpage

\end{document}